\documentclass[12pt]{article}
\usepackage{epsfig}

\newcommand{\beq}{\begin{equation}}
\newcommand{\eeq}{\end{equation}}
\newcommand{\bea}{\begin{eqnarray}}
\newcommand{\eea}{\end{eqnarray}}
\newcommand{\nn}{\nonumber}
\newcommand{\rhobar}{\bar{\rho}}
\newcommand{\etabar}{\bar{\eta}}
\newcommand{\epsilonk}{\varepsilon_K}

\newcommand{\dms}{\Delta m_s}

\newcommand{\vcb}{\left | {V_{cb}} \right |}
\newcommand{\vub}{\left | {V_{ub}} \right |}

\newcommand{\mev}{{\rm MeV}}

\def\utfit{{\bf{U}}\kern-.24em{\bf{T}}\kern-.21em{\it{fit}}\@}
\def\utangles{{\bf{U}}\kern-.24em{\bf{T}}\kern-.21em{\it{angles}}\@}
\def\utlattice{{\bf{U}}\kern-.24em{\bf{T}}\kern-.21em{\it{lattice}}\@}

\def\Title#1{\begin{center} {\Large {\bf #1} } \end{center}}

\begin{document}

\Title{Lattice QCD, Flavor Physics and the Unitarity Triangle Analysis}

\begin{center}{\large \bf Contribution to the proceedings of HQL06,\\
Munich, October 16th-20th 2006}\end{center}

\bigskip\bigskip


\begin{raggedright}

{\it Vittorio Lubicz \index{Lubicz, V.}\\
Dipartimento di Fisica, Universit\`a di Roma Tre and INFN Sezione di Roma Tre\\
Via della Vasca Navale 84 \\
I-00146 Rome, ITALY}
\bigskip\bigskip
\end{raggedright}

\begin{abstract}
Lattice QCD has always played a relevant role in the studies of flavor physics
and, in particular, in the Unitarity Triangle (UT) analysis. Before the starting
of the B factories, this analysis relied on the results of lattice QCD
simulations to relate the experimental determinations of semileptonic B decays,
$K-\bar K$ and $B_{d,s}-\bar B_{d,s}$ mixing to the CKM parameters. In the last
years much more information has been obtained from the direct determination of
the UT angles from non-leptonic B decays. In this talk, after a presentation of
recent averages of lattice QCD results, we compare the outcome of the
``classical" UT analysis (\utlattice) with the analysis based on the angles
determinations (\utangles). We discuss the role of the different determinations
of $V_{ub}$, and show that current data do not favour the value measured in
inclusive decays. Finally we show that the recent measurement of $\Delta m_s$,
combined with $\Delta m_d$ and $\epsilonk$, allows a quite accurate extraction
of the values of the hadronic parameters, $\hat B_K$, $f_{Bs} \, \hat
B^{1/2}_{Bs}$ and $\xi$. These values, obtained ``experimentally'' by assuming
the validity of the Standard Model, are compared with the theoretical
predictions from lattice QCD.
\vspace{1pc}
\end{abstract}

\section{Introduction}
Lattice QCD has always played a relevant role in the history of the UT fit since
the very beginning. At the time when the B factories had not started yet, the
``classical'' UT analysis relied on the results of lattice QCD simulations to
relate the experimental studies of semileptonic $B$ decays, $B_{d,s}-\bar
B_{d,s}$ and $K-\bar K$ mixing to the CKM parameters. Despite the
lattice results were mostly obtained in the quenched approximation at that time,
and some of the experimental determinations were still rather rough, these
analyses allowed to reach at least three important results for flavor physics
in the Standard Model (SM): i) the amount of indirect CP violation observed in
kaon mixing ($\epsilonk$) was shown to be fully consistent with the expectation
based on the CKM mechanism of CP violation; ii) and iii) quite accurate
predictions for $\sin 2 \beta$ and $\Delta m_s$ were obtained.

\begin{figure}[htb]
\begin{center}
\includegraphics[width=7.0cm]{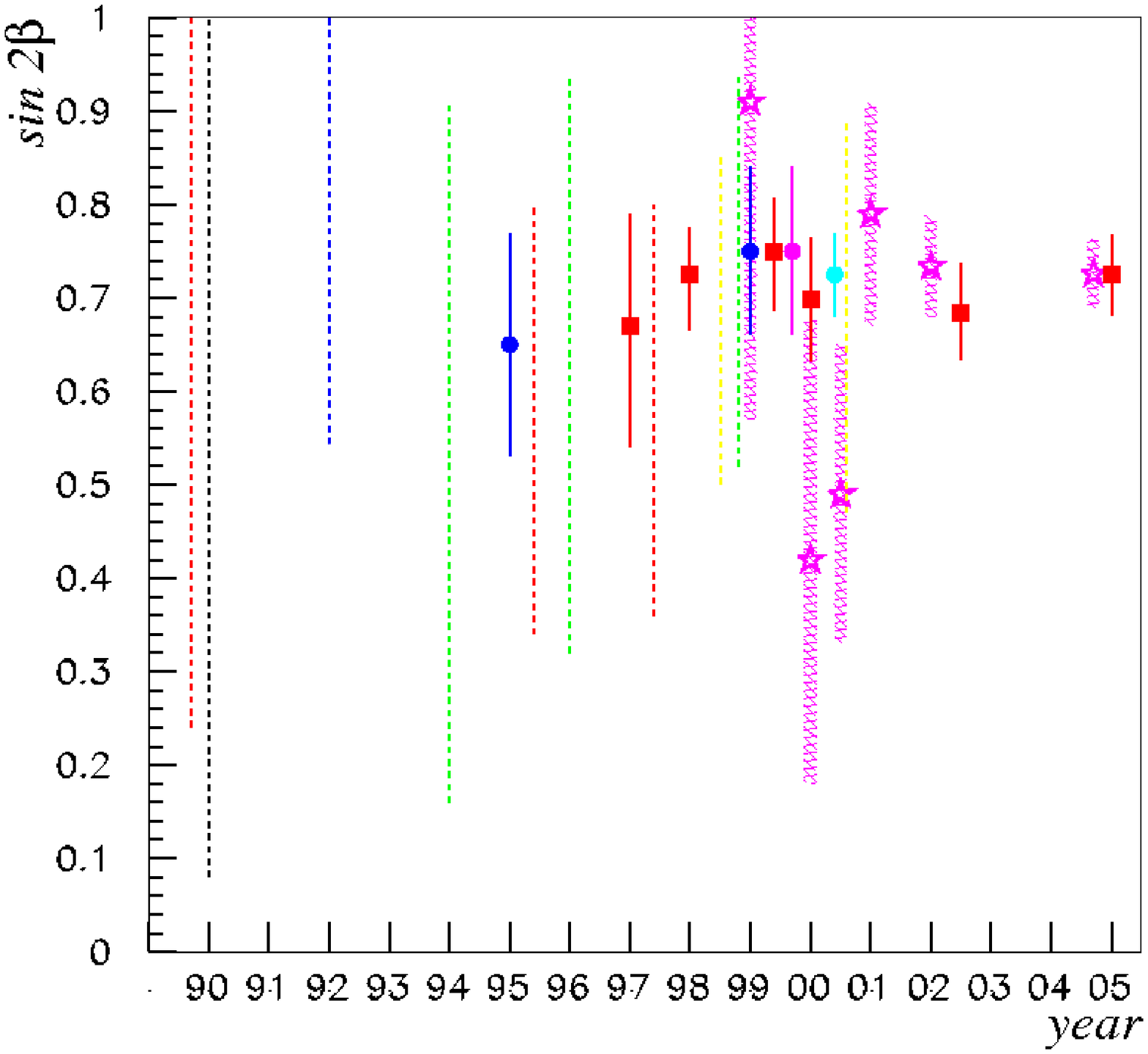}
\includegraphics[width=7.0cm]{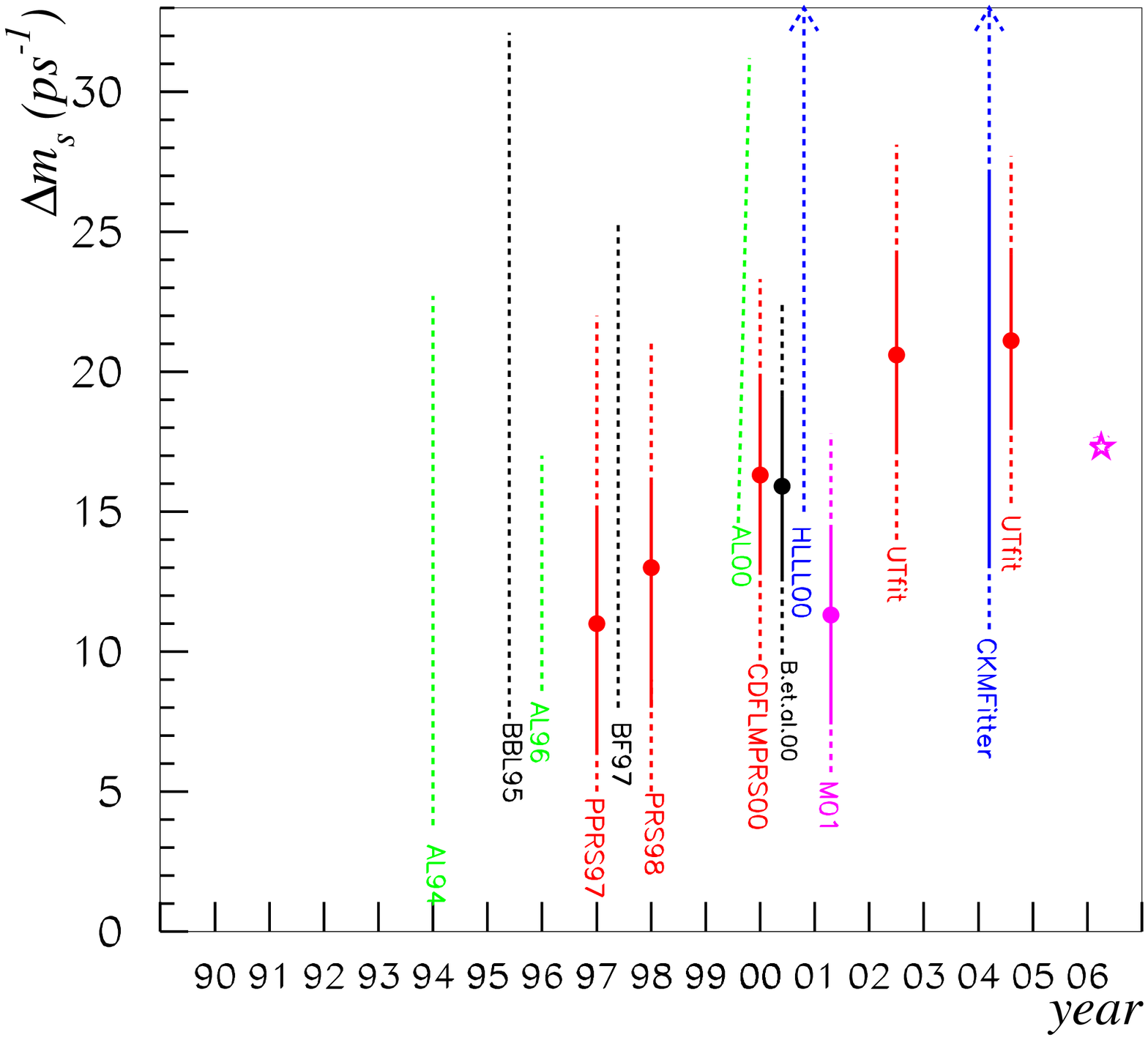}
\caption{\sl Evolution of the predictions for $\sin 2\beta$ (left) and $\Delta
m_s$ (right) over the years. The corresponding experimental determinations are
indicated by larger bands and stars. See \cite{UTfit,istant} for the full list
of original references}.
\label{fig:storie}
\end{center}
\end{figure}

Predictions of $\sin 2\beta$ exist since more than 15 years, see
Fig.~\ref{fig:storie} (left): the first indication of a large value of this
parameter, namely $\sin 2\beta > 0.55$, dates back to
1992~\cite{Lusignoli:1991bm}. In 1995, the prediction $\sin 2\beta= 0.65 \pm
0.12$ was derived~\cite{prei1}. Five years later, when direct measurements were
not available yet, we obtained the more accurate estimate $\sin 2\beta= 0.698
\pm 0.066$~\cite{utfitseminal}. These results are in remarkable agreement with
the present experimental average, $\sin 2\beta= 0.675 \pm
0.026$~\cite{sin2b_exp}.

\begin{figure}[htb]
\begin{center}
\includegraphics[width=7.0cm]{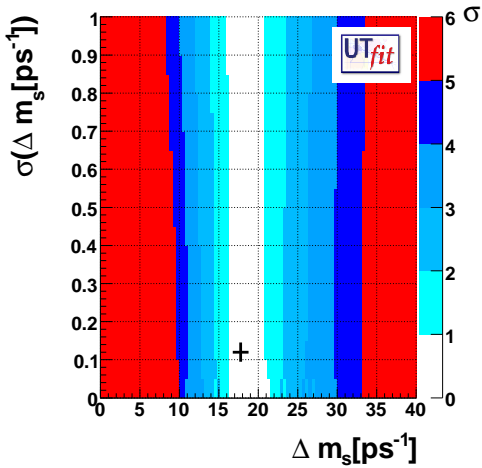}
\caption{\sl Compatibility plot of the measured value $\Delta m_s= (17.77 \pm
0.12)\,\,\mathrm{ps}^{-1}$ with the SM expectation from the other constraints
of the UT fit.}
\label{fig:pull1}
\end{center}
\end{figure}

A similar situation holds for $\Delta m_s$. A precise indirect determination
of $\Delta m_s$ from the other constraints of the UT fit was available since
1997: $\Delta m_s \in [6.5,15.0]$~ps$^{-1}$ at $68\%$ probability and $\Delta
m_s<22$~ps$^{-1}$ at $95\%$ probability~\cite{paganini}. A compilation of the
predictions for $\Delta m_{s}$ by various collaborations as a function of time
is shown in Fig.~\ref{fig:storie} (right). As can be seen from the plot, even in
recent years, and despite the improved measurements, in some
approaches~\cite{CKMfitter1,CKMfitter2} the predicted range was very large (or
corresponds only to a lower bound~\cite{CKMfitter1}). An upgraded version of our
SM ``prediction'' for $\Delta m_s$ obtained from the full UT fit is $\Delta m_s=
(18.4 \pm 2.4)$~ps$^{-1}$~\cite{UTfit}, in remarkable agreement with the direct
measurement $\Delta m_s = (17.77 \pm 0.12) \,\,\mathrm{ps}^{-1}$~\cite{dms_exp}.
In Fig.~\ref{fig:pull1} we show the compatibility plot for $\Delta m_s$, which
illustrates the agreement, at better than $1\, \sigma$ level, of the measured
value with the SM expectation.

In the last years, we got much more information on the UT from the direct
determinations of the angles $\alpha$, $\beta$ and $\gamma$, obtained at the B
factories from the studies of CP asymmetries in non-leptonic B decays. In the
following we will call the ensemble of these measurements \utangles: they allow
a determination of $\rhobar$ and $\etabar$ independently of the hadronic
parameters computed on the lattice. The precision in constraining $\rhobar$ and
$\etabar$ from the \utangles\ is by now comparable to that obtained from the
lattice-related constraints, denoted as \utlattice. The latter include the
determination of $\vub /\vcb$ from semileptonic B decays, $\epsilon_K$, $\Delta
m_d$ and $\Delta m_s$.

In this talk, after a presentation of recent averages of lattice QCD results, we
will compare the outcome of the \utlattice\ and \utangles\ analyses. We will
discuss the role played by the different determinations of $V_{ub}$ and show
that the value measured in inclusive decays is not favoured by the data.
Finally, we will show that the recent measurement of the $B_s$-meson mixing
amplitude $\Delta m_s$, combined with $\Delta m_d$ and $\epsilon_K$, allows a
quite accurate extraction of the values of the hadronic matrix elements relevant
for $K-\bar K$ and $B_{s,d}-\bar B_{s,d}$ mixing. Assuming the validity of the
SM, we determine these hadronic quantities from the experimental data and
compare them with recent lattice calculations. The content of this talk is
mostly based on Ref.~\cite{istant}, but we take here the opportunity to update
the results by taking into account the most recent experimental
findings~\cite{UTfit}.

\section{Averages of Lattice QCD results}
Lattice QCD is the theoretical tool of choice to compute hadronic quantities.
Being only based on first principles, it does not introduce additional free
parameters besides the fundamental couplings of QCD, namely the strong coupling
constant and the quark masses. In addition, the systematic uncertainties
affecting the results of lattice calculations can be systematically reduced in
time, with the continuously increasing availability of computing power and the
development of new algorithms and improved theoretical techniques.

In spite of the appealing features of the lattice approach, the accuracy
currently reached in the determination of the hadronic matrix elements is
typically still at the level of 10-15\%. So far, the main limiting factor for
achieving an improved precision has been the lack of sufficient computing power,
which has often prevented the possibility of performing ``full QCD" simulations
and forced the introduction of the quenched approximation. In this approximation
an error is introduced which, besides being process dependent, is also difficult
to reliably estimate.

Most of the lattice calculations relevant to $B$-physics used so far the
quenched approximation. There are few exception in which $N_f=2$ dynamical
quarks are included in the QCD vacuum
fluctuations~\cite{AliKhan:2001jg}-\cite{Bernard:2002pc} and only a single
calculation with $N_f=2+1$~\cite{Gray:2005ad}~\footnote{The ``+1" indicates that
an heavier strange quark is included in the sea, besides the two degenerate up
and down quarks}. A similar situation holds for the lattice studies of $K-\bar
K$ mixing. Only three unquenched calculations of the kaon parameter $B_K$ have
been produced so far, one including $N_f=2$ dynamical quarks~\cite{Aoki:2004ht}
and two with $N_f=2+1$~\cite{Gamiz:2006sq,Antonio:2007pb}.

It is important to emphasise that the quenched results, in spite of being
unsatisfactory for having been obtained with unrealistic $N_f=0$, have the
advantage that the whole methodology of extracting the desired information from
the simulation has been developed and understood, the procedure of
non-perturbative renormalization has been implemented, and the whole plethora of
results have been checked by many different groups, using various versions of
the gauge and fermionic lattice actions. In a number of cases even the continuum
extrapolation has been shown to be smooth. The unquenched studies, on the other
hand, are sound for being unquenched (although the dynamical quarks are still
much heavier than the physical up and down quarks). However, the consequences of
the so called fourth-root trick implemented with the staggered fermion
formulation are not clear, the non-perturbative renormalization in most of the
cases is not carried out, and the results have not been checked yet by different
groups. In this respect, unquenching is still a work in progress. At the same
time, it is should be noted that the capability of decreasing significantly the
values of the simulated light quark masses in recent unquenched calculations has
allowed to largely reduce the uncertainty associated with the chiral
extrapolation. Particularly relevant cases, in this respect, are the
determinations of the pseudoscalar decay constant $f_B$ and of the ratio
$\xi$~\cite{Gray:2005ad}. For all these reasons it is important to take into
account, when producing averages of lattice QCD results, the outcome of the
several more recent lattice calculations. The average lattice values that are
being used in the UT analysis, for the quantities relevant to $K$- and
$B$-physics, are:
\bea
&& \hat B_K =  0.79 \pm 0.04 \pm 0.08 \ \ \cite{dawson} \ , \nn \\
&& f_{Bs} \, \hat B^{1/2}_{Bs} = 262 \pm 35 ~\mev \quad , \quad 
\xi =1.23 \pm 0.06 \ , \nn \\
&& f_{Bd} = 189 \pm 27 ~\mev \quad , \quad f_{Bs} = 230 \pm 30 ~\mev  \ \
\cite{hashimoto} \ .
\label{eq:averages}
\eea

\section{\utlattice, \utangles\ and role of $\vub$}
\begin{figure}[htb]
\begin{center}
\includegraphics[width=4.9cm]{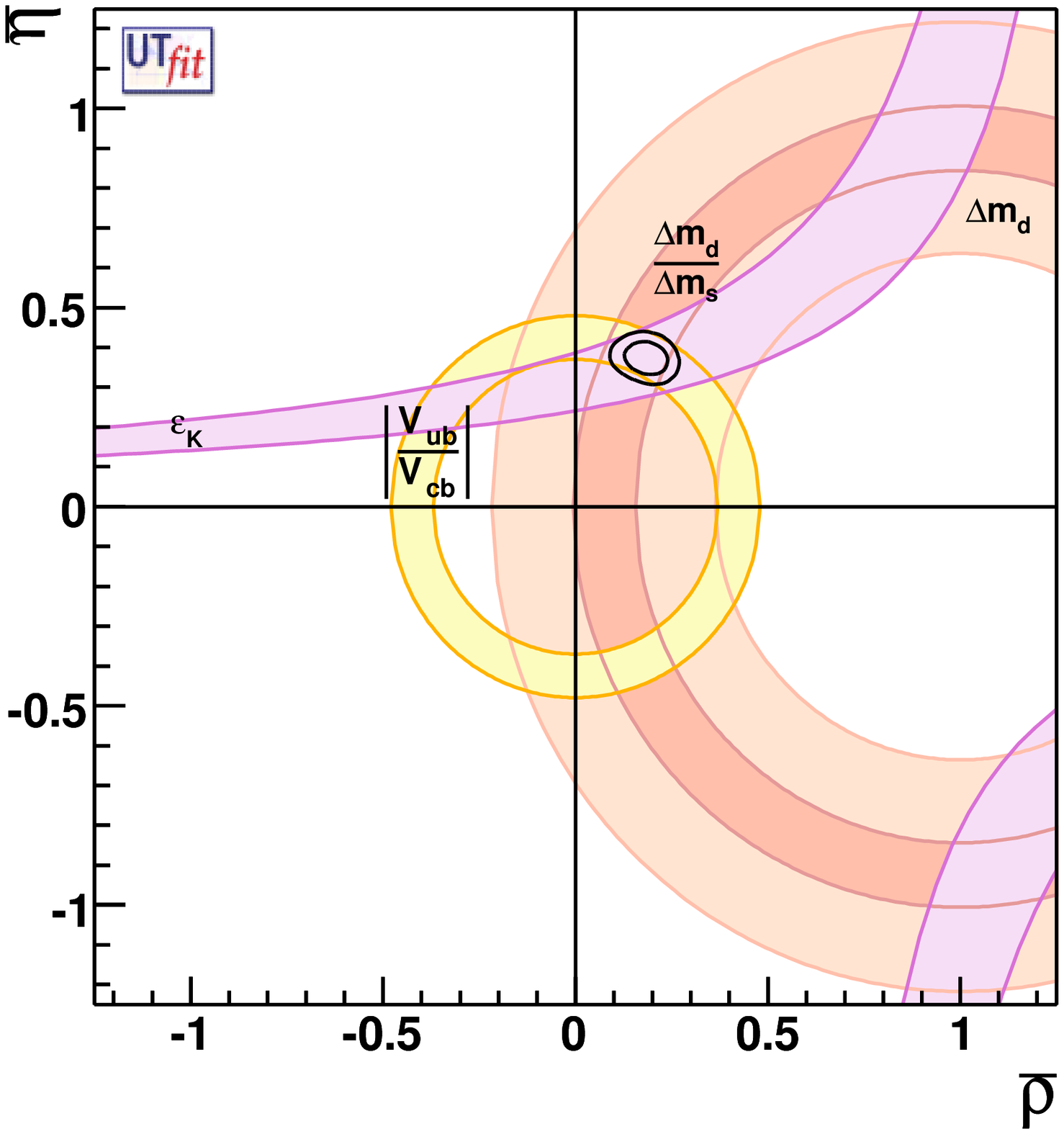}
\includegraphics[width=4.9cm]{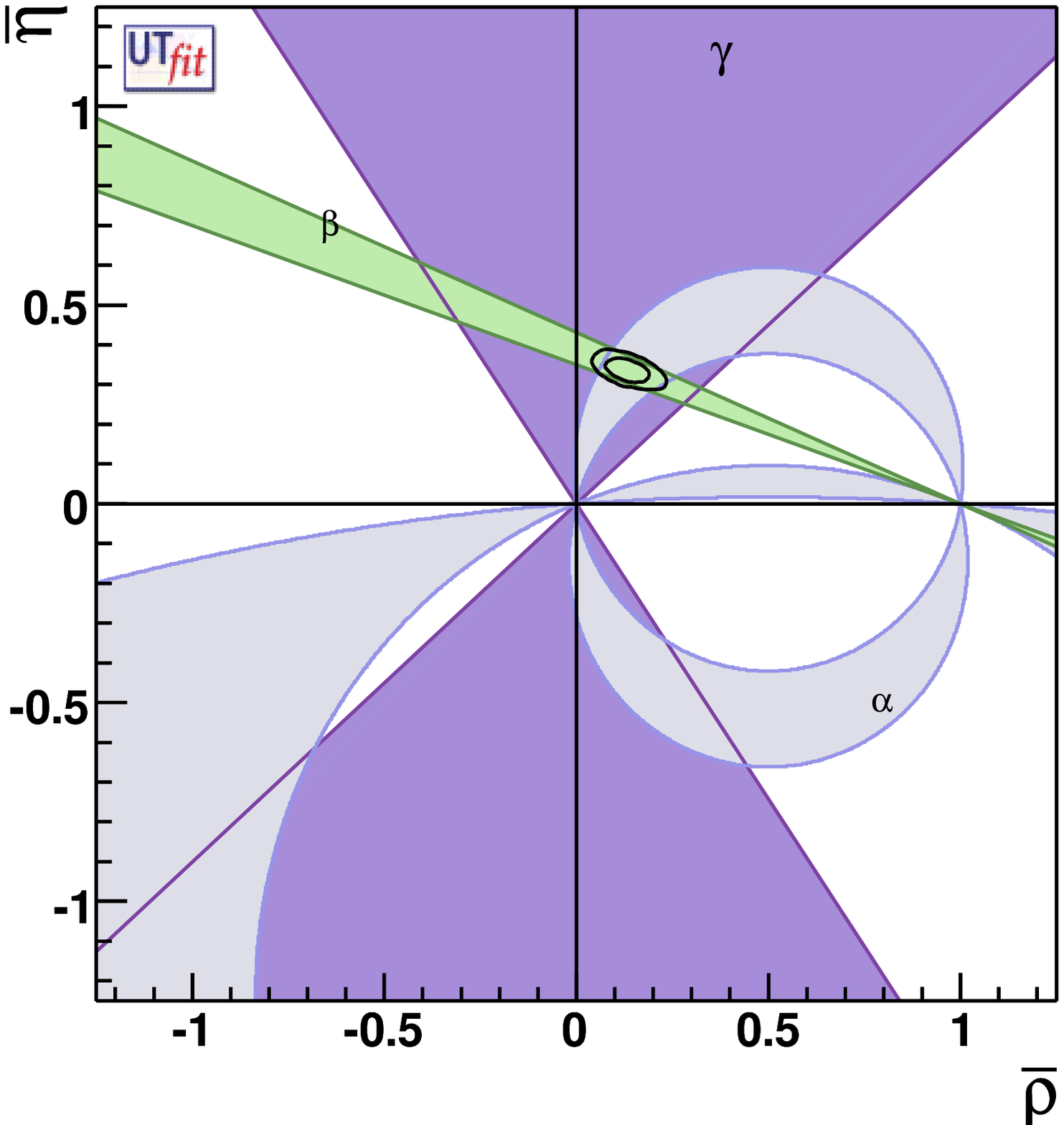}
\includegraphics[width=4.9cm]{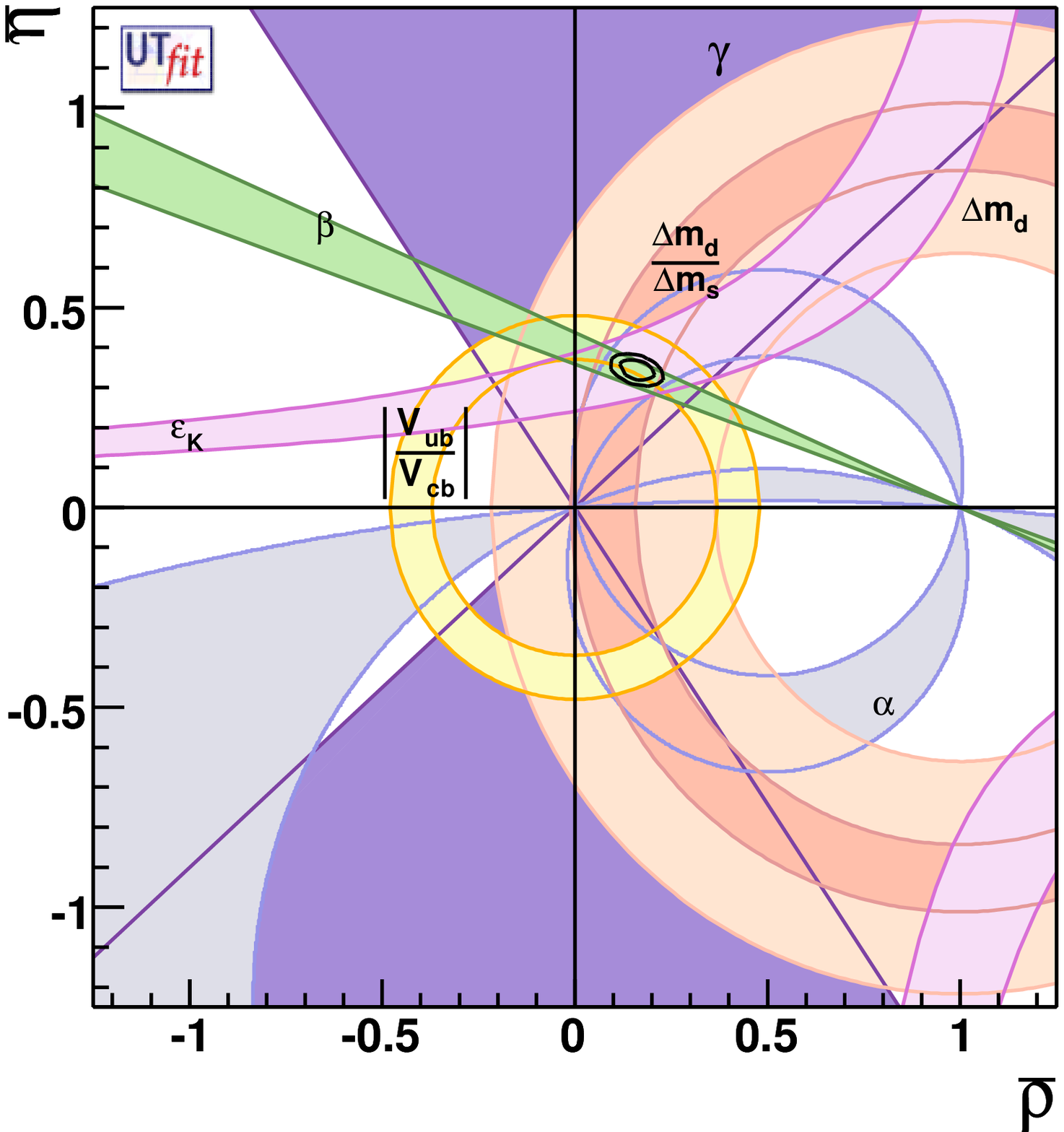}
\caption{\sl Determination of $\rhobar$ and $\etabar$ from \utlattice\ (left),
\utangles\ (center) and the full UT (right) analyses. The $68\%$ and $95\%$
total probability contours are shown, together with the $95\%$ probability
regions from the individual constraints.}
\label{fig:CKM_fit_today}
\end{center}
\end{figure}

In Fig.~\ref{fig:CKM_fit_today} we show the results of the UT fit as obtained
from the lattice-related constraints, \utlattice, the direct determinations of
the UT angles, \utangles, and the full analysis~\cite{UTfit,istant}.
The corresponding determinations of $\rhobar$ and $\etabar$, as derived
independently from the \utlattice\ and \utangles\ analyses, are
\bea
& {\bf UT}lattice & \qquad \quad  {\bf UT}angles  \nn \\ 
& \rhobar = 0.188 \pm 0.036&,  \quad \rhobar = 0.134 \pm 0.039 \nn \\ 
& \etabar = 0.371 \pm 0.027 &,  \quad \etabar = 0.335 \pm 0.020 ~.
\label{rhoeta}
\eea
We firstly note that the errors have comparable sizes, i.e. the two analyses
have reached at present a comparable level of accuracy. It is also interesting
to observe that the \utangles\ fit, based on the direct determination of the UT
angles, does not rely at all on theoretical calculation of the hadronic matrix
elements, for which there was a long debate about the treatment of values and
error distributions~\cite{yellowb}. In the \utangles\ analysis, the treatment of
theoretical errors is not an issue.

\begin{figure}[htb]
\begin{center}
\includegraphics[width=7.0cm]{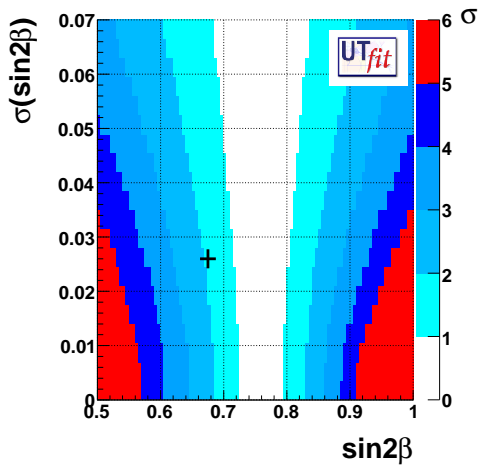}
\caption{\sl Compatibility plot of the experimental value of $\sin2\beta$ and
the prediction from the rest of the UT fit.\label{fig:sin2bpull}}
\end{center}
\end{figure}

The results in eq.~(\ref{rhoeta}) also show the existence of a tension between
the values of $\rhobar$ and $\etabar$ obtained from the two analyses. This is
also illustrated by the effects of the various constraints in the full UT fit,
shown in the right plot of Fig.~\ref{fig:CKM_fit_today}. It mainly appears to be
a tension between the (presently quite accurate) measurement of $\sin 2\beta$
and the constraint coming from the determination of $\vert V_{ub}\vert$ from
semileptonic B decays. The poor agreement is also evidenced by the comparison
between the experimental value $\sin 2\beta=0.675(26)$ and the value obtained
by using all the other constraints in the UT fit, i.e. $\sin 2\beta=0.759(37)$.
The compatibility plot of $\sin 2\beta$, presented in Fig.~\ref{fig:sin2bpull},
shows that this tension is indeed at the 2$\,\sigma$ level.

In order to further investigate where the tension comes from, it is worth
recalling that there is a systematic difference between the exclusive and
inclusive determination of $\vert V_{ub}\vert$ (the inclusive values are
always larger than the exclusive ones). Current averages and errors are:
\bea
\label{vubexp}
\vert V_{ub} \vert^{\rm{excl.}} = (35.0 \pm 4.0) \times 10^{-4} \nn \\
\vert V_{ub} \vert^{\rm{incl.}} = (44.9 \pm 3.3) \times 10^{-4}
\eea
These determinations also rely on non-perturbative hadronic quantities: the
semileptonic form factors for exclusive semileptonic B decays and the HQET
parameters $\bar \Lambda$, $\lambda_1$ and $\lambda_2$, for inclusive ones.
While the form factors are determined from lattice QCD calculations and QCD
sum rules, the HQET parameters are extracted, together with $\vub$, directly
the from the fits of the experimental data. In this latter case, however, a
certain amount of model dependence has to be introduced, and various
approaches (BLNP~\cite{blnp}, DGE~\cite{dge}, BLL~\cite{bll}) are currently
considered. The systematic difference between the exclusive and inclusive
determination of $\vub$ might be explained by the uncertainties of the
theoretical approaches.

\begin{figure}[htb]
\begin{center}
\includegraphics[width=7.0cm]{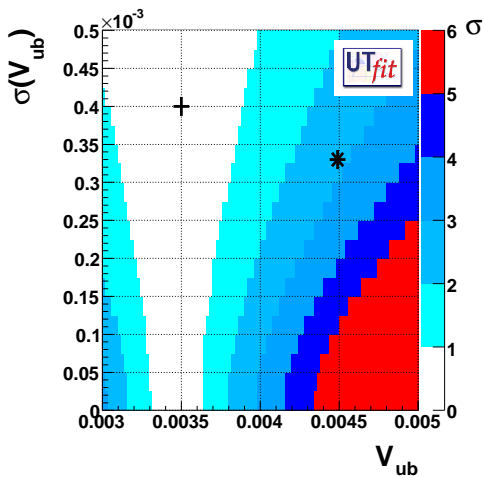}
\caption{\sl Compatibility plot between the direct determinations of
$\vert V_{ub}\vert$ from exclusive (cross) and inclusive (star) analyses
and the rest of the UT fit.}
\label{fig:pull_vub}
\end{center}
\end{figure}

In Fig.~\ref{fig:pull_vub} we show the compatibility plot between the direct
determinations of $\vub$ from exclusive and inclusive analyses and the rest
of the fit.
While the exclusive determination is in remarkable agreement with the
expectation from the UT analysis in the SM, the inclusive value of $\vub$ shows
a deviation which is almost at the $3\, \sigma$ level. Our analysis suggests
that, although both the exclusive and inclusive results are still
compatible, there could be some problem with the theoretical calculations,
and/or with the estimate of the uncertainties, of inclusive $b \to u $
semileptonic decays. In order to clearly solve this tension, an effort
should be made to increase the precision of lattice QCD determinations of
the form factors of $B \to \pi$ and $B\to \rho$ semileptonic decays,
providing all of them in the unquenched case, with low light quark masses
and studying the continuum limit of the relevant form factors. Note that
this tension among exclusive and inclusive calculations is a peculiarity of
$\vub$, since the inclusive and exclusive determinations of $\vcb$ are in
much better agreement.

\section{Constraints on lattice parameters}
Assuming the validity of the SM, the constraints in the $\rhobar$-$\etabar$
plane from \utangles\ and semileptonic $B$ decays, combined with the
measurements of $\Delta m_d$, $\Delta m_s$ and $\epsilon_K$, allow the
``experimental'' determination of several hadronic quantities which were
previously taken from lattice QCD calculations. This approach has the
advantage that the extracted values of $\hat B_K$ and of the B mixing
parameters $f_{Bs,d} \, \hat B^{1/2}_{Bs,d}$ (or equivalently $f_{Bs} \,
\hat B^{1/2}_{Bs}$ and $\xi$) can be compared directly with the theoretical
predictions from lattice QCD.

Besides $\hat B_K$, $f_{Bs}\, \hat B^{1/2}_{Bs}$ and $\xi$, we can also
extract the values of the leptonic decay constants $f_{B}$ and $f_{Bs}$ from
the fit, using in addition the lattice values of the B mixing parameters,
$\hat B_{Bd}=\hat B_{Bs}=1.28 \pm 0.05 \pm 0.09$~\cite{hashimoto}. With
respect to the decay constants, the lattice calculations of the B-parameters
are typically affected by smaller uncertainties, since some sources of
systematic errors (like those related to the determination of the lattice
scale and to chiral logs effects) are either absent or largely reduced.
Moreover, quenched and unquenched estimates of the B parameters are found to
be quite consistent within each others.

\begin{figure}[htb]
\begin{center}
\includegraphics[width=4.9cm]{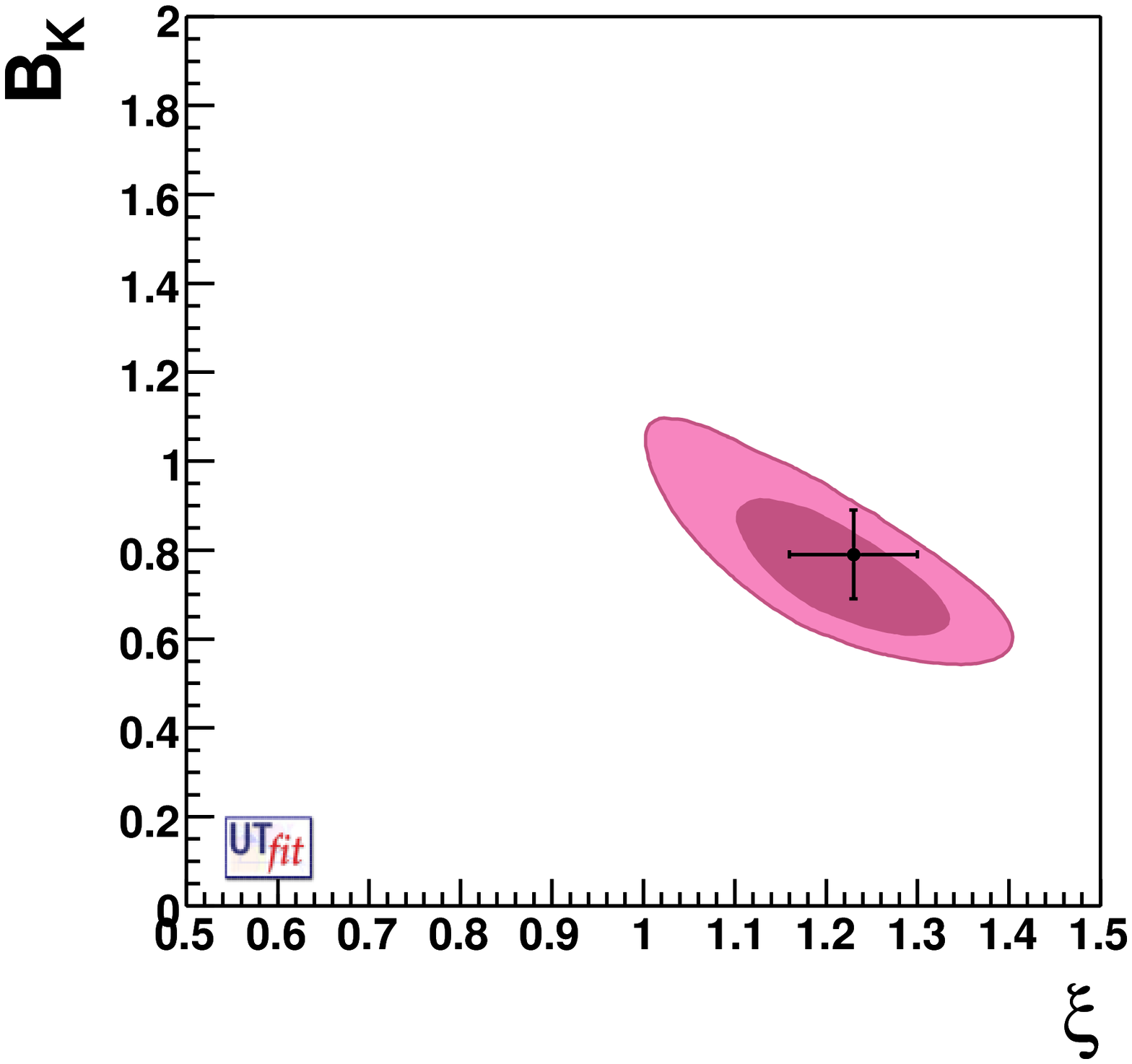}
\includegraphics[width=4.9cm]{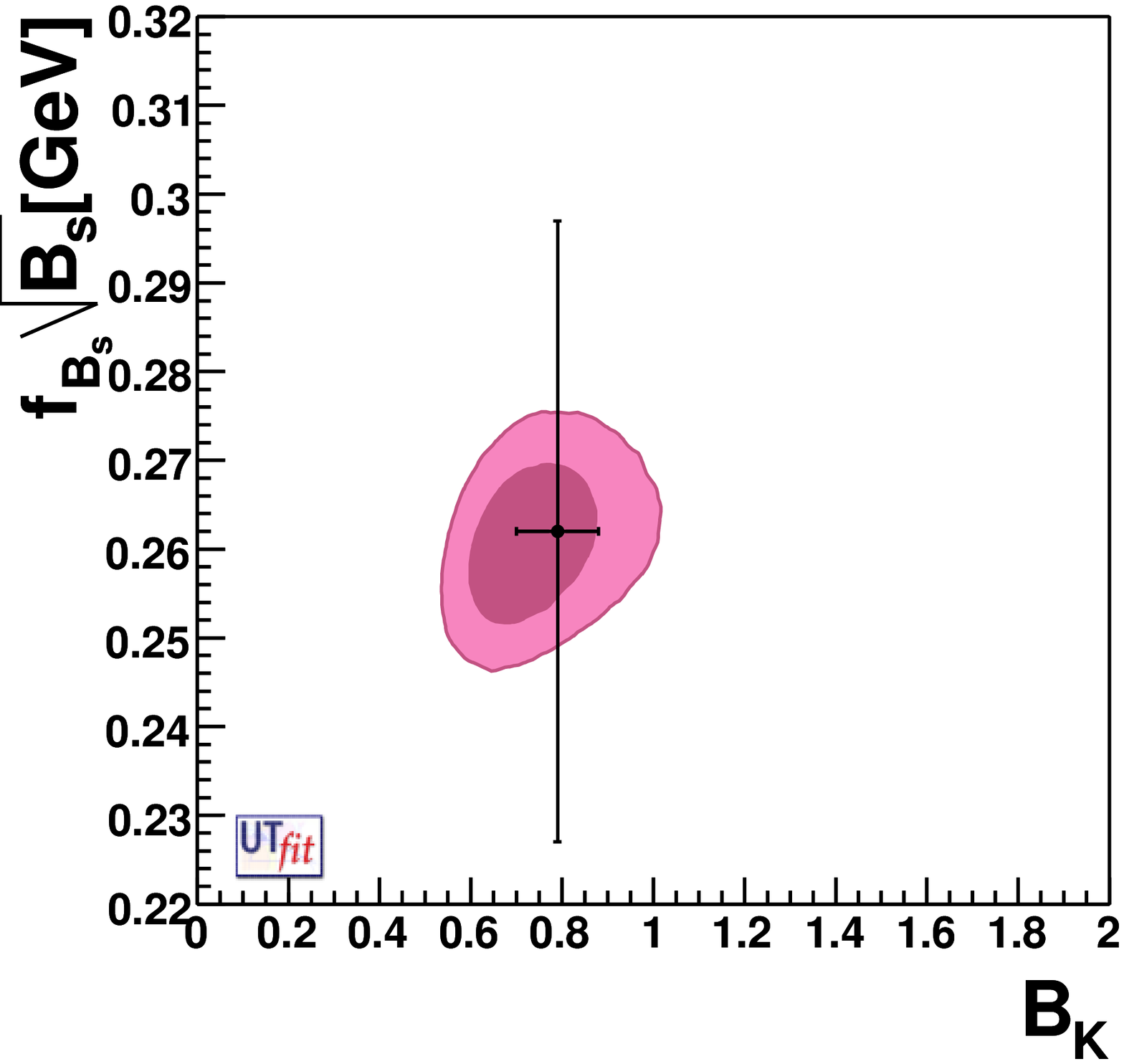}
\includegraphics[width=4.9cm]{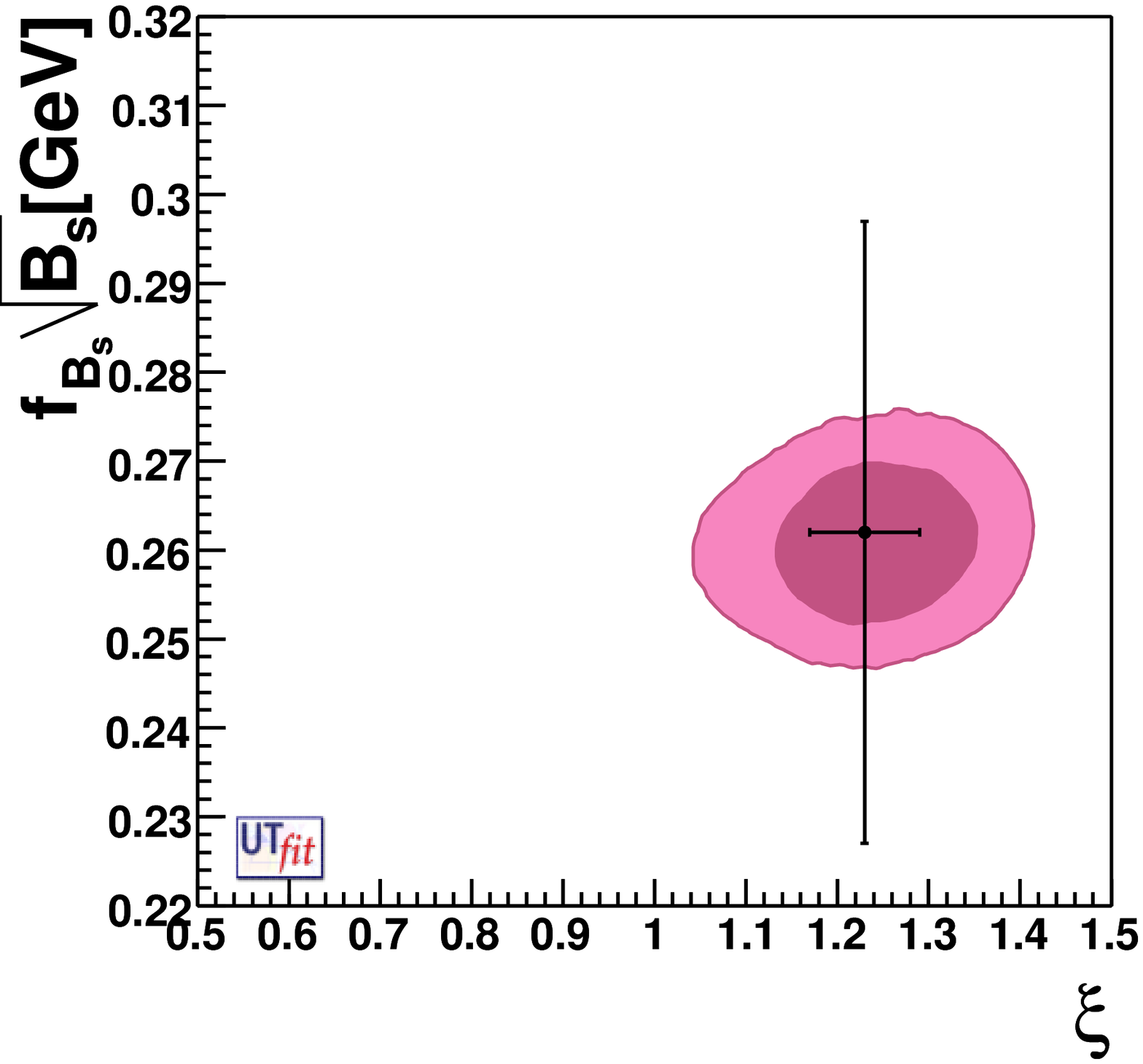}
\caption{\sl Two-dimensional constraints in the $B_K$ vs. $\xi$, $f_{Bs}
\sqrt{\hat B_s}$ vs. $B_K$ and $f_{Bs}\sqrt{\hat B_s}$ vs. $\xi$ planes,
using the \utangles\ result for the CKM matrix and the experimental
information on $\epsilonk$, $\Delta m_d$ and $\Delta m_s$. The error bars
show the results from lattice QCD calculations.
\label{fig:2dlattice}}
\end{center}
\end{figure}

The results for $\hat B_K$, $f_{Bs}\, \hat B^{1/2}_{Bs}$, $\xi$, $f_{B}$ and
$f_{Bs}$ extracted from the UT fit are presented in Tab.~\ref{tab:hadronic},
together with the average values from lattice QCD
calculations~\cite{dawson,hashimoto} also given in eq.~(\ref{eq:averages}).
\begin{table}[b]
\begin{center}
\begin{tabular}{|cccccc|}
\hline
& $\hat B_K$ &  $f_{Bs} \, \hat B^{1/2}_{Bs}$ (MeV) &  $\xi$ &
$f_{Bd}$ (MeV) & $f_{Bs}$ (MeV) \\ \hline
UT fit  & $0.75 \pm 0.09$ & $261 \pm 6$ &  $1.24 \pm 0.08$ &  $187 \pm 13$ &
$231 \pm 9$ \\
LQCD  & $0.79 \pm 0.04 \pm 0.08$ & $262 \pm 35$ &  $1.23 \pm 0.06$ &
$189 \pm 27$ & $230 \pm 30$ \\
\hline
\end{tabular} 
\caption{\sl Comparison between determinations of the hadronic parameters from
the UT fit and the averages from lattice QCD calculations
(eq.~(\ref{eq:averages})).}
\label{tab:hadronic}
\end{center}
\end{table}
The agreement between the two determinations is remarkable. On the one hand,
this comparison provides additional evidence of the spectacular success of
the SM in describing flavor physics. On the other hand, the results given
in Tab.~\ref{tab:hadronic} illustrate the accuracy and the reliability
reached at present by lattice QCD calculations. The allowed probability
regions in the $B_K$ vs. $\xi$, $f_{Bs} \hat B^{1/2}_{Bs}$ vs. $B_K$ and
$f_{Bs}\hat B^{1/2}_{Bs}$ vs. $\xi$ planes derived from the UT fit are shown
in Fig.~\ref{fig:2dlattice}, together with the results from lattice QCD
calculations.

The UTfit determination of $f_{Bs}\, \hat B^{1/2}_{Bs}$ has an accuracy of
about 2\%, which reflects the precision reached in the experimental
determination of $\dms$. This uncertainty on $f_{Bs} \, \hat B^{1/2}_{Bs}$
is by far smaller than the one reached by lattice QCD calculations. On the
other hand, the errors on the UTfit and lattice QCD determinations of $\hat
B_K$ and $\xi$ have comparable size, leaving in these cases the opportunity
for further improvement of the theoretical calculations.

It is worth recalling that the phenomenological extraction of the hadronic
parameters and the comparison with lattice results assumes the validity of the
SM and it is meaningful in this framework only. A similar strategy could be
followed in any given extension of the SM when enough experimental information
is available. In general, however, a model-independent UT analysis beyond the SM
cannot be carried out without some ``a priori'' theoretical knowledge of the
relevant hadronic parameters. For this reason, the error in the calculation of
the hadronic matrix elements affects the uncertainties in the determination of
the New Physics parameters~\cite{UTNP,UTNP2}, which is at present one of the
main motivations for the studies of flavor physics.

\section{Conclusions}
The recent precise determination of $\Delta m_s$ by the CDF Collaboration allows
a substantial improvement of the accuracy of the UT fit. Thanks to this
measurement, it is possible to extract from the experiments the value of the
relevant hadronic parameters, assuming the validity of the SM. The results of
this fit turn out to be in remarkable agreement with the theoretical predictions
of lattice QCD. It is also remarkable that the measurement of $\Delta m_s$,
combined with all the information coming from the UT fit, allows the
determination of $f_{Bs} \, \hat B^{1/2}_{Bs}$ with an accuracy of about 2\%
($f_{Bs}\, \hat B^{1/2}_{Bs}= 261\pm6$~MeV).

The only exception to the general consistency of the UT fit is given by the
inclusive semileptonic $b \to u$ decays, the analysis of which relies on the
parameters of the shape function and other model-dependent assumptions. We
observed that the present determination of $\vert V_{ub}\vert$ using inclusive
methods is disfavoured by all other constraints almost at the 3$\,\sigma$ level.
This can come either from the fact that the central value of $\vert V_{ub}\vert$
from inclusive decays is too large, or from the smallness of the estimated
error, or both. We think that it is worth investigating whether the theoretical
uncertainty of the inclusive analysis has been realistically estimated. At the
same time, an effort should be done for a substantial improvement of the
theoretical and experimental accuracy in the extraction of $\vert V_{ub}\vert$
from exclusive decays. Indeed in the future a confirmation of these results,
with smaller errors, might reveal the presence of New Physics in the generalised
UT analysis~\cite{UTNP2}.

\bigskip
I warmly thank the organisers for the very pleasant and stimulating atmosphere
of the conference. I am indebted to all my friends of the \utfit\
Collaboration, with which most of the results presented in this talk have been
obtained.

\end{document}